# Deep Learning-based Segmentation of Cerebral Aneurysms in 3D TOF-MRA using Coarse-to-Fine Framework

M. Chen, C. Geng, D.-D. Wang, J.-J. Zhang, R.-Y. Di, F.-M. Li, Z.-Y. Zhou, S.-R. Piao, Y.-X. Li, Y.-K. Dai

## ABSTRACT

**BACKGROUND AND PURPOSE:** Cerebral aneurysm is one of the most common cerebrovascular diseases, and SAH caused by its rupture has a very high mortality and disability rate. Existing automatic segmentation methods based on DLMs with TOF-MRA modality could not segment edge voxels very well, so that our goal is to realize more accurate segmentation of cerebral aneurysms in 3D TOF-MRA with the help of DLMs.

**MATERIALS AND METHODS:** In this research, we proposed an automatic segmentation framework of cerebral aneurysm in 3D TOF-MRA. The framework was composed of two segmentation networks ranging from coarse to fine. The coarse segmentation network, namely DeepMedic, completed the coarse segmentation of cerebral aneurysms, and the processed results were fed into the fine segmentation network, namely dual-channel SE_3D U-Net trained with weighted loss function, for fine segmentation. Images from ADAM2020 (n=113) were used for training and validation and images from another center (n=45) were used for testing. The segmentation metrics we used include DSC, HD, and VS.

**RESULTS:** The trained cerebral aneurysm segmentation model achieved DSC of 0.75, HD of 1.52, and VS of 0.91 on validation cohort. On the totally independent test cohort, our method achieved the highest DSC of 0.12, the lowest HD of 11.61, and the highest VS of 0.16 in comparison with state-of-the-art segmentation networks.

**CONCLUSIONS:** The coarse-to-fine framework, which composed of DeepMedic and dual-channel SE_3D U-Net can segment cerebral aneurysms in 3D TOF-MRA with a superior accuracy.

**ABBREVIATIONS:** DLM = Deep Learning Model; ADAM = Aneurysm Detection And segMentation Challenge; HD = Hausdorff Distance; VS = Volumetric Similarity

Cerebral aneurysm(CA) is a sort of tumor-like protrusion of the arterial wall caused by the abnormal expansion of the lumen of the cerebral artery. CAs mostly occur in the circle of Willis and are common in people aged between 40 and 60. Rupture of CAs is the main cause of SAH[1]. However, the early missed diagnosis rate of CAs is high, there is no warnings for rupture, and the mortality and disability rate of the first rupture is as high as about 30%[2]. At the same time, it will induce different degrees of cerebral vasospasm and acute hydrocephalus. Therefore, early detection of unruptured

From Xuzhou Medical University (M.C., Y.-K.D.), Xuzhou, China; From Suzhou Institute of Biomedical Engineering and Technology (C.G., Z.-Y.Z., F.-M.L., Y.-K.D.), Chinese Academy of Sciences, Suzhou, China; From Department of Radiology (D.-D.W., R.-Y.D., S.-R.P., Y.-X.L.), Huashan Hospital, Fudan University, Shanghai, China; From Suzhou University of Science and Technology (J.-J.Z.), Suzhou, China; From Jinan Guoke Medical Engineering Technology Development co., LTD(Z.-Y.Z.), Jinan, China.

Y.-K. Dai and Y.-X. Li are both corresponding authors.

Please address correspondence to Yakang Dai, PhD, Xuzhou Medical University, 209 Tongshan road, Xuzhou 221000, China; e-mail: daiyk@sibet.ac.cn; Yuxin Li, MD, Department of Radiology, Huashan Hospital, Fudan University, 12 Wulumuqi Middle Road, Shanghai 200040, China, e-mail: liyuxin76@126.com

cerebral aneurysms(UCAs) and designing of treatment plans can save the lives of patients. TOF-MRA is one of the most commonly used screening methods for CAs in outpatient service and physical examination.

It is important to determine the size, shape, and location of the CA[3] when making a treatment plan. Despite the continuous development of angiography, the manual measurement of CAs is still a time-consuming and laborious task due to the different shapes and sizes and their location on complex intracranial vessels. In order to solve the challenges encountered in aneurysm segmentation, researchers have proposed a variety of technical methods based on computer-aided diagnosis. At present, there are few researches on aneurysm segmentation algorithm in 3D TOF-MRA. Only Sichtermann[4] et al. used convolutional neural network(CNN) to perform the detection task on 3D TOF-MRA dataset for the first time, and DSC reached 0.53; In terms of other image modalities, Shahzad[5] et al. proposed an ensemble learning method of multiple models, their model achieved DSC of 0.81 on CTA datasets. Jin[6] et al. designed a network of encoding-decoding structure, and added an improved ConvLSTM[7] to the network by utilizing the timing characteristics of DSA images, reaching a DSC of 0.53.

In this paper, we proposed a coarse-to-fine framework to segment cerebral aneurysms in 3D TOF-MRA, which is composed of DeepMedic[8] and dual-channel SE_3D U-Net. And we trained the dual-channel SE_3D U-Net with a novel weighted loss function to focus on edge voxels that were difficult to segment accurately in segmentation task.

## MATERIALS AND METHODS

### Dataset

The subjects in our study were divided into 2 groups: 78 subjects in the training cohort and 45 subjects in the external test cohort. Figure 1 shows displays of CAs. The training cohort from ADAM 2020[9] included 78 subjects with 113 images, of which 20 were normal people, each containing 1.93 UCAs on average. The test cohort from Huashan Hospital Affiliated to Fudan University included 45 subjects with 45 images, each containing 1.02 UCAs on average. The following imaging parameters were used for dataset: training cohort - Repetition Time/Echo Time, 22.0msecs/3.45msec; Magnetic Field Strength, 3.0T; test cohort - Repetition Time/Echo Time, 25msecs/5.7msec; Magnetic Field Strength, 3.0T. During preprocessing, we performed the following operations: automatic cerebral artery extraction, Z-Score normalization[10], vascular contour extraction. During extracting the vascular contour, a three-dimensional isotropic operator was used to slightly dilate the artery extraction result to smooth the surface of vessels, and then vascular contour was extracted by the sobel[11] operator as one channel of the network input. For the training cohort, did the following further processing:

Since our goal was to segment UCAs, we did not take the ruptured aneurysms into consideration in our work. The originally labeled ruptured aneurysms (label=2) were processed as background (label=0). At the same time, under the coarse-to-fine segmentation framework, the tasks of DeepMedic and dual-channel SE_3D U-Net were different, so the annotation maps of training cohort were different for the networks. When training DeepMedic, we adaptively dilated the aneurysm (label=1) according to the size; and then, as the input of SE_3D U-Net was the VOI cropped according to the coarse segmentation result, we took the center of the aneurysm as the center of the cube, and cropped the VOI of 64 x 64 x 64. For normal cases without CAs, we cropped the VOI of the same size with the center point of background as the center. The processed dataset was divided into training cohort and validation cohort at a ratio of 4:1. The training cohort was augmented to eight times through flipping, discrete Gaussian filtering[12], and histogram equalization.

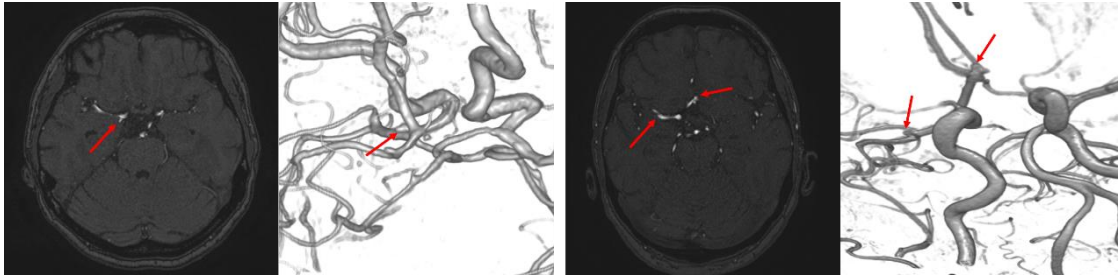

FIG 1. Displays of single and multiple CAs images in 3D Slicer[13]. The left group is images of single CA, and the right group is images of double CAs.

**Methods**

In this study, we adopted a coarse-to-fine framework for segmentation. The preprocessed data was first passed through the coarse segmentation network, which was DeepMedic, to detect the aneurysm area, namely VOI, and then cropped the vessel image and the vascular contour image at the same position according to the VOI coordinates, and feed it into the fine segmentation network, which was dual-channel SE_3D U-Net trained with weighted loss function we proposed. Finally, the output of the network was restored to the corresponding position and size of the original image by resampling, so as to obtain the segmentation of aneurysm. The algorithm framework for testing is shown in Figure 2.

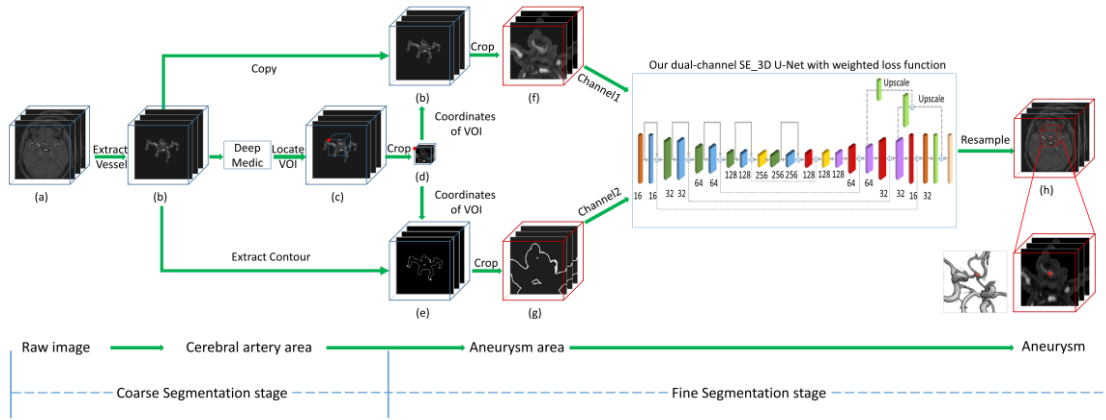

FIG 2. Overview of framework for testing: (a)3D TOF-MRA image; (b)Cerebral artery extraction from (a); (c)CA segmented by DeepMedic in coarse segmentation stage and the red point means coordinates and size of VOI; (d)VOI of (c) that cropped according to the red point; (e) Vascular contour extracted from(b) by sobel operator; (f) VOI of (b) that cropped according to the red point; (g) VOI of (e) that cropped according to the red point; (h) The resample prediction of fine segmentation network that superimposed in (a).

**Coarse segmentation stage**

DeepMedic is a commonly used CNN in medical image segmentation tasks [4, 5], which is composed of two parallel branches: the high-resolution branch and the low-resolution branch. The output of the low-resolution branch is resampled to the size of high-resolution branch's output, and then add the resampled output to the latter. The edge of final segmentation is improved through 3D Fully Connected Conditional Random Field [14] (FC-CRF).

When training DeepMedic, we performed five-fold cross-validation on a Tesla V100 (NIVIDA) GPU with 16-GB VRAM, and the training time was 25-42 hours. The main software environment included: Python 3.6, CUDA 10.0, and Tensorflow-gpu 1.14.0. In this article, the following parameters were set: the number of iterations was 700; the batch size was 10; the learning rate was 1e-3 initially and dropped gradually; L1 was regularized to 1e-6, L2 was regularized to 1e-4; RmsProp[15] was used as an optimizer.

**Fine segmentation stage**

Figure 3 shows our proposed dual-channel SE_3D U-Net. The network is optimized based on the variant 3D U-Net proposed by Isensee[16]. Similar to U-Net, the network is still a four-layer deep structure. In the encoding path, except for the input layer is a 3 x 3 x 3 convolution with stride 1, each layer is consisted of a 3 x 3 x 3 convolution with stride 2 followed by a context block. The context block is composed of a 3 x 3 x 3 convolution with stride 2 followed by a dropout layer with a dropout probability of 0.3. Additionally, residual connections are embedded between the convolution block and the context block to reduce the loss of feature maps. After the penultimate context block, the SE[17] block is embedded in it. As a spatial attention mechanism, the SE block can assign weights to effective feature channels and suppress invalid features. In the decoding path, except that the output layer is 3 x 3 x 3 convolution with stride 1, the other layers consist of a localization block and an up-sampling block. After the first up-sampling block, we also embed the SE block. The localization block includes a 3 x 3 x 3 convolution with stride 1 and a 1 x 1 x 1 convolution, which input is the summation that concatenates output of the upsampling block and the context block. Meanwhile, segmentation layers were employed as deep supervision at different layers in decoding pathway. The final output is obtained by adding outputs of segmentation layers and activating with softmax.

The input of SE_3D U-Net is dual-channel data composed of vessel image and its contour image. It is well known that vascular regions where CAs are located on are morphologically different from the normal ones. Therefore, the vascular contour is extracted, which is input into the network as prior knowledge to guide model learning.

When training dual-channel SE_3D U-Net, we performed five-fold cross-validation on a GeForce RTX 2080 Ti (NIVIDA) GPU with 11-GB VRAM, and the training time was 4-6 hours. The main software environment included: Python 3.6, CUDA 10.0, Keras 2.3.1, and Tensorflow-gpu 2.0.0. In this article, the following parameters were set: the number of iterations was 500; the batch size was 1; the learning rate was 5e-4 initially and decreased to 1/2 of the previous time if the validation loss did not improve within 10 iterations, training would be stopped after 50 epochs without the validation loss improving; Adam was used as an optimizer.

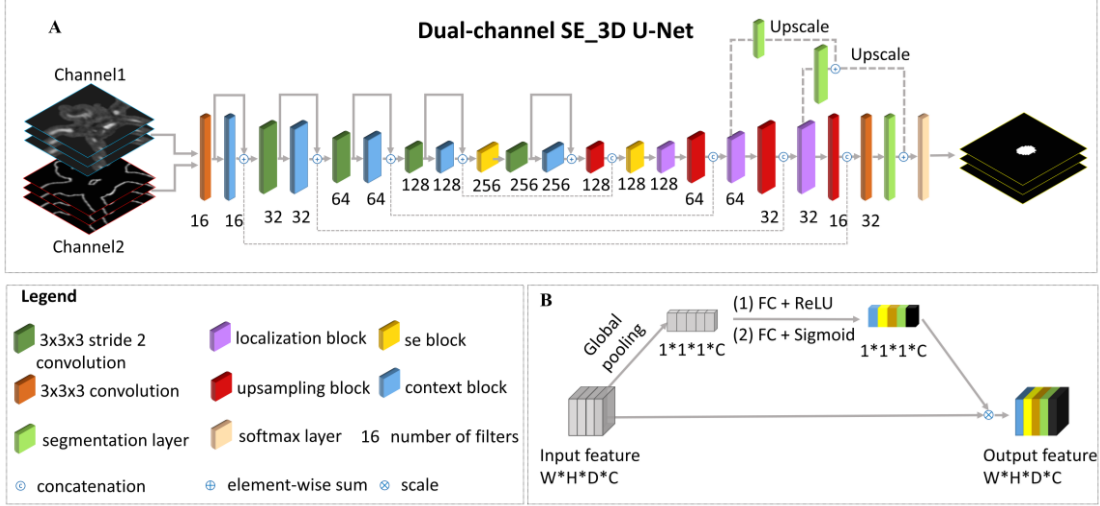

FIG 3. A, The full architecture of dual-channel SE_3D U-Net, which input is vessel images and vascular contour images. B, The architecture of 3D SE block that embed in A, and various colors of channels mean different weights.

**Loss function**

In this work, inspired by focal loss [18], we designed weighted dice loss (WDL). The purpose was that when the network learned some edge voxels that were difficult to segment, the loss value became larger, so as to promote the network to focus on the edge voxels. In detail, during training process, the overlap between label and prediction was evaluated by DSC. When the overlap was too little, it proved that the segmentation performance was poor. At this time, the loss value was weighted, that is to say, the worse the segmentation performance was, the higher the loss value would be, or vice versa. GT is the abbreviation of ground truth, Pred refers to prediction, β and S are average constant terms, β is derived from experimental inference, and S is an empirical value of 0.0001. The formula of our loss function is as shown in (1) below:

$$WDL = (1-dsc)^\beta \left( -2 \times \frac{GT \cdot Pred + \frac{S}{2}}{GT^2 + Pred^2 + S} \right) \quad (1)$$

In order to satisfy the conditions raised above, we analyzed the value interval of β. Since WDL is obtained by multiplying two factor terms, and the latter is always negative, the monotonicity of the former determines the monotonicity of the whole function. We set $G(x,y) = -(1-x)^y$, and hoped that in a certain interval of $y$, when $x$ was smaller, $G(x,y)$ became larger; when $x$ was larger, $G(x,y)$ became smaller. By simulating the changing trend of function and comparing different values of $y$ in Figure 4, we found that when $y$ took a negative value, $G(x,y)$ did not show a significant change; when $y$ took a positive value, $G(x,y)$ would show obvious fluctuations with the change of the value interval of $y$, especially in the interval of [0,1], the monotonicity of $G(x,y)$ met the expectations raised above. Therefore, it could be concluded that when the value interval of β was [0,1], the function of weighting poorly segmented voxels could be achieved.

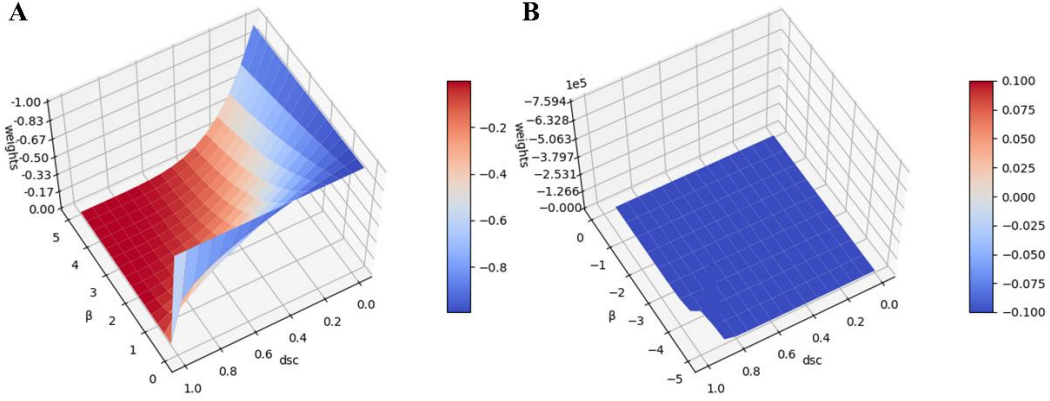

FIG 4. The simulating function diagram with β in different value intervals. The closer to the red, the larger the function value, and the closer to the blue, the smaller the function value. The toolkit for drawing images comes from the matplotlib of python. A, When β is a positive number, the change trend of function G. B, When β is a negative number, the change trend of function G.

**Evaluation**

For DeepMedic, the following metrics are used: Sensitivity, False Positive count and Positive Predicted Value (PPV), where TP indicates the number of regions that are correctly predicted as aneurysms, and FN indicates the number of areas that were mistakenly predicted as vessels. FP means the number of vessel areas that were mistakenly predicted as aneurysms. The metrics are defined as (2), (3), (4) shown:

$$Sensitivity = \frac{TP}{TP + FN} \quad (2)$$

$$Fasle\ Positive\ count = \frac{FP}{Number\ of\ subjects} \quad (3)$$

$$PPV = \frac{TP}{TP + FP} \quad (4)$$

For our dual-channel SE_3D U-Net and the overall segmentation framework, the following metrics are used: DSC, HD, and VS, where GT is the abbreviation for ground truth, and Pred refers to prediction. The formulas of these metrics are shown in (5), (6), (7):

$$DSC = 2 \times \frac{GT \cap Pred}{GT \cup Pred} \quad (5)$$

$$HD = \max\bigl(h(GT, Pred), h(Pred, GT)\bigr) \quad (6)$$

$$VS = 1 - \frac{|GT - Pred|}{GT + Pred} \quad (7)$$

## RESULTS

As mentioned above, DeepMedic and dual-channel SE_3D U-Net were separately trained in this study, so we evaluated DeepMedic, dual-channel SE_3D U-Net and the whole algorithm framework respectively. When comparing the results, we referred to the ranking method proposed by Timmins[19] to calculate the average ranking value of metrics. The model with the lowest score was considered as the best one.

The training cohort (n=90), which was partitioned from dataset (n=113), was augmented for DeepMedic training and validation. In this paper, we performed five-fold cross-validation, and selected the best model as the final coarse segmentation network. The comparison of metrics shows that the best model reached sensitivity of 100.00%, false positive count of 0.99FPs/case, and PPV of 53.02.

Table 1: The five-fold cross-validation prediction of DeepMedic

|  | Fold_0 | Fold_1 | Fold_2 | Fold_3 | Fold_4 |
|---|---|---|---|---|---|
| Sensitivity(%) | 98.51 | 100.00 | 100.00 | **100.00** | 100.00 |
| FP count(FPs/case) | **0.82** | 1.30 | 1.27 | 0.99 | 1.59 |
| PPV | **57.35** | 46.22 | 46.76 | 53.02 | 41.22 |
| rank | 0.3333 | 0.4378 | 0.4137 | **0.1631** | 0.6667 |

The dataset for dual-channel SE_3D U-Net was VOIs (n=147) cropped from dataset (n=113). After the same data augmentation as operated in training DeepMedic, it was put into dual-channel SE_3D U-Net for training and validation. According to the preceding analysis, $\beta$ of WDL should fall in [0,1]. In order to get the best value of $\beta$, we referred to the value of the exponential term in focal loss[18], and compared the experimental results based on three metrics. It can be seen from Table 2 that when $\beta$ is 1, the model reached DSC of 0.75, HD of 27.06, and VS of 0.64, which had the best effect on improving network performance.

Table 2: Comparison between various value of $\beta$ of weighted dice loss function

| $\beta$ | 0 | 0.1 | 0.3 | 0.5 | 0.7 | 0.9 | 1 |
|---|---|---|---|---|---|---|---|
| DSC | 0.77±0.18 | 0.76±0.16 | **0.82±0.13** | 0.71±0.13 | 0.80±0.15 | 0.76±0.15 | 0.75±0.15 |
| HD | 36.41±6.40 | 36.25±6.85 | 36.17±6.93 | 35.33±6.85 | 34.05±7.60 | 30.66±11.45 | **27.06±13.51** |
| VS | 0.45±0.30 | 0.43±0.30 | 0.41±0.30 | 0.38±0.32 | 0.48±0.32 | 0.62±0.32 | **0.64±0.35** |
| rank | 0.7284 | 0.7787 | 0.6196 | 0.9615 | 0.5149 | 0.3358 | **0.2121** |

Through ablation experiments on validation cohort, the effects of dual-channel input (DCI) and WDL on the performance of SE_3D U-Net were compared, as shown in Table 3. The baseline network in the table represents single-channel SE_3D U-Net without training with WDL. Through experiments, it could be found that when the vascular contour is added as one channel of network input, DSC was increased by 0.03, HD was reduced by 3.51, and VS was increased by 0.25; when the weighted loss function was used to train the model, DSC was increased by 0.03, HD was reduced by 0.28, and VS was increased by 0.02. Compared with the baseline model, DSC was increased by 0.08, HD was decreased by 2.73, and VS was increased by 0.17, which proved that the performance of our proposed network was better.

Table 3: Comparison between baseline network, DCI + baseline network, baseline network trained with WDL and network we proposed

|  | Baseline | DCI + Baseline | WDL + Baseline | Proposed |
|---|---|---|---|---|
| DSC | 0.73±0.16 | 0.76±0.19 | 0.79±0.17 | **0.81±0.17** |
| HD | 36.27±6.88 | 32.76±11.38 | 35.99±6.50 | **17.93±13.91** |
| VS | 0.42±0.30 | 0.67±0.27 | 0.44±0.31 | **0.84±0.21** |
| rank | 1.0 | 0.6128 | 0.7290 | **0.0** |

Table 4 shows performance of the dual-channel SE_3D U-Net trained with WDL on five-fold cross-validation. The best model was selected as the fine segmentation network and was connected in series with DeepMedic as the final framework for aneurysm segmentation.

Table 4: Results of five-fold cross-validation in dual-channel SE_3D U-Net trained with WDL

|  | Fold_0 | Fold_1 | Fold_2 | Fold_3 | Fold_4 |
|---|---|---|---|---|---|
| DSC | 0.81±0.17 | 0.79±0.15 | 0.81±0.16 | **0.82±0.15** | 0.80±0.16 |
| HD | **17.93±13.91** | 25.42±12.49 | 22.61±13.49 | 19.64±14.57 | 21.44±13.13 |
| VS | 0.84±0.21 | 0.78±0.22 | 0.80±0.24 | **0.85±0.18** | 0.82±0.22 |
| rank | 0.1587 | 1.0 | 0.5574 | **0.0761** | 0.5213 |

In order to visually display the segmentation of the network proposed in this paper, two TOF-MRA images were randomly selected for manual labeling, namely golden standard. The segmentation obtained through baseline network and our proposed network were both on the bias of DeepMedic as coarse segmentation network in Figure 5. Comparing segmentation of the baseline network with that of our proposed network in Figure 5, we could find that the segmentation of proposed method was closer to that of manual annotation. The baseline method cannot precisely segment the boundary of region, while the method we proposed was sensitive to the boundary, which segmentation was more detailed.

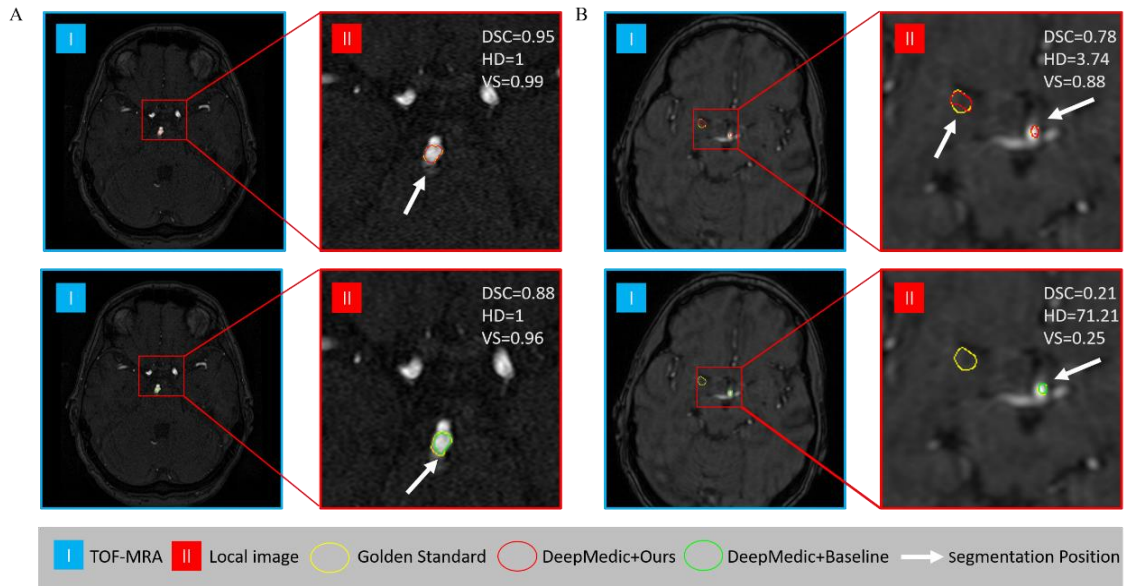

FIG 5. A, Single CA segmentation through manual annotation, DeepMedic + baseline network and DeepMedic + network proposed, although the baseline method can segment the aneurysm, its boundary segmentation accuracy was not as good as the method proposed in this paper. B, Multiple CAs segmentation through manual annotation, DeepMedic + baseline network and DeepMedic + network proposed, it could be seen that on the basis of the same coarse segmentation network, there was omissions in the segmentation of baseline network.

The dataset for testing is an independent cohort from another center, and was not used for training in our work. Meanwhile, the imaging parameters and scanners were also different from training cohort. Table 5 shows the segmentation results obtained through a variety of fine segmentation network on the external test cohort, including baseline network, our proposed method, and four state-of-the-art segmentation networks under the condition of using DeepMedic as the coarse segmentation network. The four state-of-the-art segmentation networks all used the training process of the fine segmentation of the baseline network in this article. The test cohort was preprocessed and fed into the overall segmentation framework, and the prediction were resampled to the original size for evaluation. It can be seen that the final DSC, HD and VS of the proposed segmentation method on test cohort reached 0.12, 11.61, and 0.16. Compared with other networks, the three metrics were significantly improved. The results in table 5 proved the robustness of our proposed method. When the training cohort and the test cohort were completely unrelated, our proposed method showed better segmentation performance than baseline network and state-of-the-art methods considering the difference in dataset.

Table 5: Results of overall framework on test cohort

|     | baseline | Ours | 3D U-Net | V-Net | CE-Net | nnUnet |
| --- | --- | --- | --- | --- | --- | --- |
| DSC | 0.07±0.16 | 0.12±0.14 | 0.00 | 0.04±0.10 | 0.06±0.10 | 0.07±0.13 |
| HD | 32.51±40.96 | 11.61±5.19 | Nan | 72.18±77.15 | 37.69±55.65 | 23.67±10.23 |
| VS | 0.12±0.23 | 0.16±0.19 | 0.00 | 0.07±0.14 | 0.13±0.13 | 0.12±0.17 |

## DISCUSSION

In this study, we optimized and trained DLMs for segmenting CA in 3D TOF-MRA, and evaluated them on an independent test cohort. The following innovations were proposed: vascular contour was extracted as prior knowledge to guide the model learning; a weighted loss function was designed to improve the segmentation accuracy of the aneurysm region boundary.

The dual-channel input in this paper was composed of the vessel image and its extracted contour image. As can be seen from the results on validation cohort, the performance of dual-channel network was significantly improved compared to the current common single-channel networks. For specific performance，DSC was increased by 0.03, HD was decreased by 3.51, and VS was increased by 0.25 compared with the original model, which showed that feature extraction of input had a significant effect on model optimization. In the above metrics, the improvement of VS was particularly significant. As mentioned above, VS can evaluate the voxels similarity between the ground truth and the prediction, the improvement of VS reflects that it is feasible to extract vascular contour as prior knowledge. Morphological information can help the model pay attention to abnormal vessels with growing aneurysm during learning. In the follow-up work, we will try to extract more characteristic information, such as the surface curvature of vessels, blood flow direction, etc., to achieve generalized multi-channel input.

While verifying the exponential value of the weighted loss function, we chose the value within the interval [0,1] for experimental comparison, and when β=1, the network performance improved the most. Compared with the baseline network, the model trained with WDL increased DSC by 0.06, decreased HD by 0.28, and increased VS by 0.02. However, compared with the dual-channel network, the model failed to show a particularly significant improvement in HD and VS. When designing the weighted loss function, we aimed to help DLMs focus on the edge voxels of aneurysms during training. The voxel overlap between the ground truth and the prediction could be intuitively reflected through DSC. The improvement of DSC indicated that WDL did play a role in prompting the network to focus on the edge voxels of aneurysms during the training process. When the DCI and WDL were simultaneously applied to the fine segmentation network, the segmentation performance was significantly improved compared with the baseline network.

We used a test cohort that is completely independent of training cohort to evaluate the segmentation performance of model. In some papers with good DSC, we found the common ground was that test cohort still had similar data to training cohort. In this work, our training cohort is public dataset from ADAM2020, and the test cohort comes from the Huashan Hospital affiliated to Fudan University. Under different regions and ethnicity, collection method of TOF-MRA data is also different. The final results showed that DSC, HD and VS obtained from our proposed method are as high as 0.12, 11.61 and 0.16 respectively. In case of huge differences between data, our model showed better performance than the state-of-the-art methods. We will also transfer our algorithm proposed in this article to datasets of other lesions to further improve the robustness of the method.

## CONCLUSIONS

In this work, we proposed a coarse-to-fine segmentation framework consisting of DeepMedic and dual-channel SE_3D U-Net trained with weighted loss function. After training and validation on 113 cases of data, this study proved that our deep learning model can accurately segment cerebral aneurysms in 3D TOF-MRA. On the 45 external test cohort, the network obtained an average DSC of 0.12, HD of 11.61, and VS of 0.16.

## ACKNOWLEDGMENT

This work was supported by National Natural Science Foundation of China [81971685], National Key Technology Research Development Program [2018YFA0703101], Science and Technology Commission of Shanghai Municipality [19411951200], Youth Innovation Promotion Association CAS [2021324], Quancheng 5150 Project.